\documentclass[twocolumn,showpacs,preprintnumbers,amsmath,amssymb,showkeys]{revtex4}
\usepackage{graphicx}
\usepackage{dcolumn}
\usepackage{bm}

\begin{document}
\title{In-plane optical features of  the 
underdoped  La$_{2}$CuO$_4$ based compounds:
Theoretical multiband analysis}
\author{Ivan Kup\v{c}i\'{c}\footnote{Fax: +385-1-4680-336, \\
E-mail address: kupcic@phy.hr
(I. Kup\v{c}i\'{c})}} 
\affiliation{ Department of Physics, 
        Faculty of Science, POB 331,  HR-10 002 Zagreb, Croatia}

\begin{abstract}
The three-component $ab$-plane optical conductivity of the high-$T_c$ cuprates
is derived using the gauge invariant 
response theory, and compared to the data previously obtained 
from the optical reflectivity measurements in the
La$_{2}$CuO$_4$ based families.
	The valence electrons are described by the Emery three-band
model with the antiferromagnetic correlations 
represented by an effective  single-particle potential.
	In the $0 < \delta < 0.3$ doping range,
it is shown that the total spectral weight of the three-band model
is shared between the intra- and interband channels nearly 
in equal proportions.
	At optimum doping, the low-frequency  conductivity 
has a (non-Drude) nearly single-component form, 
which transforms with decreasing doping into 
a two-component structure.
	The mid-infrared spectral weight is found to be extremely 
sensitive to the symmetry of the effective single-particle potential,
as well as to the doping level.
	The gauge invariant form of the static and elastic Raman
vertices is determined, allowing explicit verification
of the effective mass theorem and the related conductivity sum rules.	
\end{abstract}

%



 \maketitle

\section{Introduction}
Following  the marginal Fermi liquid theory \cite{Varma},
the nested Fermi liquid theory \cite{Ruvalds}
or  the ordinary memory-function approach \cite{Opel},
the low-frequency excitations of the electronic system
involved in the optical conductivity and   electronic 
Raman scattering spectra of  the overdoped high-$T_c$ cuprates are
described in terms of the frequency-dependent effective mass $m (\omega)$ 
and the frequency-dependent relaxation time $\tau (\omega)$, 
with a minor role played by  
selection rules or by  normal-state coherence factors.
	Attempts to explain the measurements in the underdoped cuprates
using similar  single-component models have been unsuccessful.
	Anomalous low-frequency structures 
related to  the  superconducting (SC) pseudogap, the antiferromagnetic 
(AF) pseudogap, the two-magnon excitations or the density waves
are observed in the latter crystals, 
usually in  one or two (of four) ``optical'' channels
(optical conductivity and $A_{1g}$, $B_{1g}$ and $B_{2g}$
Raman symmetries).
\cite{Uchida,Cooper,Quijada,Lupi,Venturini,Sugai,Naeini,Opel}
	Any theory of the low-frequency excitations in the  underdoped 
cuprates should explain  anomalies in both optical conductivity and 
Raman scattering spectra, taking the normal-state coherence effects 
properly into account.

The previous theoretical work on the optical conductivity
of the underdoped cuprates 
\cite{Rozenberg,Dagotto,Prelovsek,Prelovsek2,Basov,Marsiglio} 
is carried out  for the one-band models with a 
particular care devoted to the treatment of strong local correlations
responsible for the AF  structure in the low-frequency conductivity,
but no attention was paid to the (interband) optical processes 
in the visible part of the spectra.
	The principal physical problem is that when such one-band
models are applied to the Raman spectra, then the resonant nature of the 
Raman scattering processes is ignored, because the intermediate 
states in the Raman scattering processes are treated in the static
approximation. \cite{Sherman2,KupcicSSC}
	The first experimental evidence	for the crucial role of the interband
processes at $\hbar \omega \approx 1.75-2.75$ eV for the low-frequency
physics of the high-$T_c$ cuprates was given by 
Cooper and coworkers \cite{Cooper},
considering the correlation between the  optical excitations across the
charge-transfer gap in YBa$_2$Cu$_3$O$_{7-x}$ and the resonant behaviour 
of the magnetic peak in the $B_{1g}$ Raman response.
	Similar (but less pronounced) dependence on the frequency of the
incident/scattered photons is also found in the Drude part of the 
Raman spectra. 
	The latter  is also accompanied by several anomalies regarding 
the relative spectral weights of three Raman polarizations.
\cite{Opel,Sugai,Naeini}

Actually, the recent interest in the Drude part of the optical conductivity and 
Raman scattering spectra of the underdoped compounds is stimulated
by the clear evidence for important role of 
the normal-state coherence effects.
	The optical conductivity and $B_{2g}$ Raman spectra scan the 
electronic states in the nodal region of the Fermi surface 
(so-called ``cold spots'' on the  Fermi surface: $k_x = k_y$), 
while the $A_{1g}$ and  $B_{1g}$ Raman scattering processes probe 
the electronic states in the vicinity of the van 
Hove points (``hot spots'').
	A striking difference between these two groups of spectra occurs
for the nearly half filled conduction band in all cases where there is a
perturbation peaked at (usually commensurate) wave vector {\bf Q} close 
to the nesting vector of the Fermi surface $2{\bf k}_{\rm F}$.
	The AF fluctuations at small dopings  
\cite{Uchida,Sugai}
and the charge-density-wave (CDW) fluctuations 
at the $1/8$ doping \cite{Venturini}
are typical examples.
	There is also a surprising  similarity between the 
mid-infrared (MIR) peak in the optical conductivity,
$\hbar \omega^{\rm ir}_{\rm peak}$, and the 
magnetic peak in the $B_{1g}$  Raman spectra.
	Both appear at nearly the same energy, exhibit similar doping
dependence and possibly have the same physical origin. 
\cite{Uchida,Cooper,Sugai,Naeini}

It is interesting also to note that  in the 
materials where the MIR peak appears at a relatively small energy,
comparable with the typical damping energy
$\Sigma \equiv \hbar /\tau$ \cite{Lupi,Venturini},
there is an essential difference between
the predictions of the usual one-band optical models and measurements.
	It is explicitly shown that the development of 
the measured MIR structure with temperature	
can be fitted well with the  orbital Kondo model
of Emery and Kivelson  \cite{Emery2}
(representing the  regime where   
$\hbar \omega^{\rm ir}_{\rm peak} \approx \Sigma $)
rather than with the mentioned  one-band models
(underdamped regime:  $\hbar \omega^{\rm ir}_{\rm peak}$ is proportional 
to characteristic correlation energy scales, such as
the Hubbard interaction $U$ \cite{Rozenberg,Dagotto} or 
the AF exchange energy $J$ \cite{Prelovsek}).

The main goal of the present analysis is the following:
	(i) to determine for the Emery three-band model \cite{Emery}
all electron-photon coupling functions (various current and Raman vertices,
to be defined below) important for the understanding of most of the 
aforementioned anomalies,
	(ii) to show that, due to the gauge invariance,
the present multi-component optical model allows a natural transformation
of the spectra from the underdamped regime into the overdamped 
regime, \cite{KupcicPB2}
	and (iii) to point out that the correlation effects peaked at
${\bf Q} \approx 2{\bf k}_{\rm F}$ 
(the AF correlations in the cuprates, or the umklapp scattering processes
in the Bechgaard salts
\cite{Schwartz,Vescoli2,Giamarchi,KupcicPB2})
affect the optical conductivity of two-dimensional (2D) and 
quasi-one-dimensional (Q1D) systems in a similar way,
leading in both cases to the dramatic decrease of the effective number 
of conduction electrons.

However, to ensure a simple analytic form of the current and Raman 
vertices, the present analysis will be limited by the use of several
approximations.
	(i) The effects of the AF correlations on the response functions 
will be included in an adiabatic way.
	(ii) The strong local correlations associated with the Hubbard term on
the copper sites will be treated in the mean-field approximation (MFA),
with the same (renormalized) copper-oxygen transfer integral describing 
the low- and high-frequency excitations.
	(iii) As usual for the in-plane features 
of the high-$T_c$ cuprates, the 2D representation of the electronic system is 
used, neglecting several important questions dependent on the
electron propagation  in the direction perpendicular to the
conduction planes, such as the 
anomalous $c$-axis conductivity or the dimensionality crossover.

In this paper, the response  of  the valence electrons in the high-$T_c$ 
cuprates to the external vector potential
is analyzed by means of the Emery three-band model \cite{Emery}.
	The  AF correlations in the nearly half filled conduction band are
modeled by an effective single-particle interaction $\Delta ({\bf k})$ 
peaked at the commensurate wave vector ${\bf Q} = (\pi/a, \pi/a )$.
	In the calculation we adopt a gauge invariant 
approach which has been proved to enable a systematic study 
of the multi-component optical conductivity spectra \cite{KupcicPB}
and which is a simple generalization of the memory-function approach
\cite{KupcicPB2}.
	It will be shown that an anisotropic-$s$ effective interaction results
in two distinct features in the low-frequency optical conductivity,
with the  relation between these two features and the
high-frequency excitations across the $pd$ charge-transfer gap
which quantitatively agrees with the experimental observations
(including the position of thresholds and the ratio between 
the respective spectral weights).

In Section 2 we describe the simplest  version of the three-band 
model which focuses on the  optical excitations across the 
$pd$ charge-transfer gap and on the low-frequency excitations across the
MIR correlation gap $ 2 \Delta ({\bf k})$
(in the rest of the text the MIR conductivity related to   
$ \Delta ({\bf k})$ will be referred to as the MIR gap structure),
and review other work.
	We  determine the gauge invariant form 
of the static and elastic Raman vertices (Appendices A and B),
which makes possible the consistent formulation of both the 
three-component optical conductivity and the related 
Raman correlation functions.
	Section 3.1 discusses the conductivity sum rules of the 
present three-component optical model, while 
in Section 3.2 the dependence of the low-frequency spectra
on the  symmetry and magnitude of the 
effective interaction $\Delta ({\bf k})$ is studied.
	Finally, the mutually competing effects of the intracell
hybridizations and the strong local correlations on the interband spectra 
of the lightly-doped conducting  planes are briefly discussed 
in Section 3.3.

\section{Theoretical model}

The electrodynamics of  valence
electrons in the high-$T_c$ cuprates is described in terms of  the extended
Hubbard model (i.e.   the Emery three-band model \cite{Emery} 
in which  all Coulomb interactions are neglected except 
for the Hubbard interaction on the copper sites $U_d$). 
	As mentioned in  Introduction,
various one-band versions of this model have been widely used to 
describe qualitatively the low-frequency optical conductivity.
	Let us first briefly review these models. 
	Then follow the detailed explanation of the present optical model
and the comparison with the experimental data.

The theoretical  study  of the underdoped regime
reveals the MIR gap structure in the optical conductivity,
in accordance with  experimental observation.  	
	In the one-band Hubbard model \cite{Rozenberg,Dagotto} 
the corresponding threshold  energy is  proportional to the Hubbard 
interaction $U$ (precisely, $\hbar \omega^{\rm ir}_{\rm peak} \approx U/2$,
with an additional, charge-transfer gap at 
$\hbar \omega^{\rm inter}_{\rm peak} \approx U$), 
while in the $t$-$J$ model \cite{Prelovsek}
this energy is related to 
the AF exchange energy $J$.
	Most of the relevant results are calculated using 
the {\it numerical} diagonalization procedure at {\it high temperatures}
(usually $T > 500$ K).
	Similarly, the anomalous (non-Drude,   single-component) 
optical conductivity measured in the overdoped compounds
can be explained in terms of the  marginal Fermi liquid theory \cite{Varma} 
or  the nested Fermi liquid theory \cite{Ruvalds}, which 
are characterized by strong quasi-particle damping effects.
	It is essential to notice that
fits of the related generalized Drude formula
to  the measured spectra require usually {\it large damping energies}
(typically $\hbar / \tau (0.5 \; {\rm eV}) \approx 0.5 \; {\rm eV}$
\cite{Uchida,Ruvalds}).

In order to treat both the local correlations related to $U_d$ and the
interplay between various small energy scales more accurately
(at low temperatures, with presumably small damping energies), as well as
to describe the relevant intracell processes explicitly, different 
three-band versions of  the Emery model are examined and  contrasted
to the measured low-  and high-frequency  optical spectra
and the related Raman spectra 
\cite{Sherman2,Kotliar,Grilli,Niksic,Kupciclong,Sherman,Mrkonjic}.
	Kotliar, Lee and Read \cite{Kotliar} 
have calculated the ground state energy of 
the electronic system self-consistently using the slave-boson
procedure.
	They have shown that  the characteristic of the
low-frequency optical processes is the renormalization of 
the copper-oxygen bond energy $t_{pd}$ 
(i.e. the copper-oxygen transfer integral, in the usual language). 
	This renormalized bond energy is found to be strongly dependent 
on doping, in particular in the lightly doped region,
where $t_{pd} \propto \sqrt{\delta}$ (hereafter, $\delta$ is the hole doping).
	In the case where $J = 0$, the MIR structure is  found to be 
related to the renormalized splitting between the oxygen and copper 
site energies, $\Delta_{pd}$  (i.e. the renormalized $pd$ dimerization 
potential) \cite{Kotliar,Grilli}.
	However, in the complete slave-boson model \cite{Grilli} 
this scale  can be associated with the 
AF exchange energy $J$ as well. 
	In both cases, the excitations across the charge-transfer
gap are connected with the bare $pd$ dimerization potential.

The electrodynamics of the underdoped cuprates may be close to 
that of the Bechgaard salts (TMTSF)$_2$X \hbox{(X = PF$_6$,} ClO$_4$).
	Namely, for the nearly half filled 2D or Q1D conduction band, 
the SC and AF (or spin-density wave) regions in the phase diagram  
are expected to be controlled by small changes in  the 
already small effective number of conduction electrons 
caused by the interplay between 
the corresponding correlation  energy and various other small energy scales.
\cite{KupcicPB2}
	The measurements on the high quality crystals 
\cite{Schwartz,Vescoli2}
have shown that the electrodynamic features of the Bechgaard salts
are strongly affected by  the umklapp scattering processes, in particular by 	
the competition between the characteristic umklapp energy
and the transfer integral in the direction perpendicular
to the highly conducting direction \cite{Giamarchi,Vescoli2}.
	In the present paper it will be proposed that the two-component 
low-frequency optical conductivity of the underdoped cuprates is connected with
a similar correlation energy  related to the AF pseudogap processes, 
and that the transformation of this two-component form
into a single-component form, in the overdoped regime, is related with 
the interplay between this scale
and the energy difference between the Fermi energy  and the van Hove energy.
	By analogy with the Bechgaard salts,
one should extend the present  
analysis by considering the competition between the characteristic AF  energy 
and the oxygen-oxygen transfer integral $t_{pp}$ 
(this is expected to be relevant for the in-plane conductivity 
in the lightly-doped regime), 
or the  interplane transfer integral $t_{\perp}$
(important for the $c$-axis conductivity and the dimensionality crossover
problem); however, these are beyond the scope of the present work.

Finally, it is important to note that 
the present optical conductivity analysis, together with the related Raman
analysis \cite{KupcicSSC}, can be easily
connected with  the Hall coefficient measurements \cite{Uchida2}, 
due particularly to the  analytic form of the intra- and interband
current and Raman vertex functions.
	It should also be noticed that
the model parameters extracted below from the optical data can be 
confirmed by angle-resolved photoemission spectroscopy (ARPES) experiments 
\cite{Ding,Liu,Ino,Lanzara,Valla,Yoshida}
(the conduction band is wide and nearly half filled),
as well as by  electric field gradient (EFG) measurements
\cite{Ishida,Ohsugi} 
(the copper-oxygen hybridization is relatively strong, again, supporting
the picture of  wide bands \cite{Kupcicefg}).

\subsection{Bare three-band Hamiltonian}
The present response theory is based on an effective 
single-particle description of the valence electrons.
	The analysis starts with
the 2D three-band  Hamiltonian 
of the form \cite{Kotliar,Kupciclong}
\begin{eqnarray}
H_0 &=& \sum_L H_0^L,
\nonumber \\
H_0^L &=& \sum_{  {\bf k} \sigma} E_L ({\bf k} )
  L^{\dagger}_{{\bf k} \sigma} L_{{\bf k} \sigma}.
\label{eq1}\end{eqnarray}
	In the hole picture,  the bonding  band (the band index $L = D$) 
is nearly half filled, while the antibonding and nonbonding bands 
($L = P$ and $ N$) are empty.

Two versions of the  model will be considered.
	The model A includes only two parameters,
the splitting between the oxygen $2p_{\sigma}$ and copper $3d_{x^2-y^2}$ 
site energies $\Delta_{pd}$
and  the average first-neighbor bond energy $t_{pd}$.
	Since  two different limits of the three-band model, 
the $U_d = 0$ limit and the MFA of  
the $U_d \rightarrow \infty$ limit (hereafter, the large $U_d $ case),
are represented by this model,
the parameters $\Delta_{pd}$ and $t_{pd}$ can be associated either 
with the bare parameters or with the parameters renormalized 
by the large $U_d $
(in the latter case, the present model is identical to the original
slave-boson model of Kotliar, Lee and Read \cite{Kotliar}).
	The distinction between these two  
situations, together with a naive phenomenological 
extension of the large $U_d$ problem,
will be briefly discussed in Section 3.3.

The model B focuses  on the bonding band only,
i.e. on an effective  low-frequency description of the conduction electrons.
	So long as the underdoped regime is in question,
the main low-frequency effects
are expected  to come from an opening of a correlation gap
in the single-particle excitation spectrum.
	These effects   are  related here to the AF correlations
and are described in the adiabatic way,
by considering the influence of 
an effective  single-particle  interaction
\begin{eqnarray}
H_{\rm corr} &=& \sum_{{\bf k} \sigma }  [ \Delta ({\bf k})
D^{\dagger}_{{\bf k}  \sigma} 
D_{{\bf k} \pm {\bf Q} \sigma} + {\rm h. c.}]
\label{eq2}\end{eqnarray} 
on the real part of the electron self-energy,
with ${\bf Q} = (\pi/a, \pi/a)$ being the AF wave vector.
	Although the effect of these correlations on the imaginary 
part of the electron self-energy is not treated explicitly,
it can be represented by a phenomenological contribution 
to the damping energies which increases linearly with frequency 
(see a comment on the generalized Drude formula given below).
	The magnitude of  $\Delta ({\bf k})$
is assumed to be real and
small, as compared with the energies $\Delta_{pd}$ and 
$t_{pd}$.

	The structure of the  Bloch functions and the 
Bloch energies of the model A  is well known \cite{Kotliar,Kupciclong}.
	On the other hand, the diagonalization of the model B 
(i.e. of the Hamiltonian 
$H_0^D + H_{\rm corr}$) is straightforward, leading to the 
dimerized bonding band with the dispersions of the upper
($L = A$)  and lower ($L = S$)  subbands given by 
\begin{eqnarray}
E_{A,S} ({\bf k}) &=&  
\frac{1}{2} [E_D ({\bf k}) + E_{\underline{D}} ({\bf k} )] 
\nonumber \\
& & 
\pm \sqrt{ \frac{1}{4} [E_D ({\bf k}) - E_{\underline{D}} ({\bf k})]^2   
+ \Delta^2({\bf k})} , 
\label{eq3}\end{eqnarray}
where $E_{\underline{D}} ({\bf k}) \equiv  E_D ({\bf k} \pm {\bf Q})$.
 	The effect of the perturbation $H_{\rm corr}$ on the Bloch functions
is given in the usual way, 
in terms of the auxiliary phase $\varphi_{\bf k}$ defined  in Appendix B.

\subsection{Coupling Hamiltonian}
According to Ref. \cite{KupcicPB}, the coupling of the 
conduction electrons to the external electromagnetic fields, 
relevant to the in-plane optical conductivity analysis, 
is given by the coupling Hamiltonian
\begin{eqnarray}
H^{\rm ext} &=& 
-\frac{1}{c} 
\sum_{LL'{\bf k} \sigma}  [A_{\alpha} ( {\bf q}_{\perp})
J^{LL'}_{\alpha} ({\bf k}) 
L^{\dagger}_{{\bf k} 
+ {\bf q}_{\perp} \sigma} 
L'_{{\bf k}  \sigma} + {\rm h. c.}] 
\nonumber   \\
&&  
 + \frac{e^2}{2 mc^2} 
\sum_{{\bf k} \sigma} [
A_{\alpha}^2 ({\bf q}_{\perp}) 
(-)\gamma^{CC} _{\alpha \alpha} ({\bf k} ;2) 
C^{\dagger}_{{\bf k} + {\bf q}_{\perp} \sigma} 
C_{{\bf k}  \sigma} 
 \nonumber \\
&& 
+ {\rm h. c.}]. 
\label{eq4}  
\end{eqnarray}
	Here $C$ is the index of the conduction band 
($C \equiv D$ in the model A and  $C \equiv A$ in the model B),
and $\alpha \in \{x, y \}$ is the photon polarization index.
	The $J^{LL'}_{\alpha} ({\bf k})$ are the coupling functions in the 
first-order term (the current vertices)
and $\gamma^{CC}_{\alpha \alpha} ({\bf k};2)$  is the coupling function 
in the bare second-order term (the bare Raman vertex).
	Furthermore, $A_{\alpha} ( {\bf q}_{\perp})$ and 
$A^2_{\alpha} ( {\bf q}_{\perp})$ are 
the Fourier transforms in space of the vector potential 
$A_{\alpha} ( {\bf r})$ and 
of its square $A^2_{\alpha} ( {\bf r})$,
respectively, and ${\bf q}_{\perp} \cdot {\bf a}_{\alpha} = 0$.
	The coupling functions of the models A and B
are shown explicitly in  Appendices A and B.

\subsection{Optical conductivity}

The doping dependence of the optical conductivity spectra
of the La$_{2-x}$Sr$_x$CuO$_4$   single crystals
has been systematically examined by Uchida et al. \cite{Uchida}.
	The underdoped crystals exhibit the behaviour which cannot be 
described as the response of a Drude metal.
	In particular, two distinct threshold energies are found 
in the underdoped regime.
	The first one (MIR threshold energy) is strongly affected by the
doping level (with the maximum in the spectra at the energy 
$\hbar \omega^{\rm ir}_{\rm peak}$ placed between 0.1 and 0.3 eV).
	In the clean (underdamped) limit of the   model B 
($\Delta_0 \gg \Sigma_3$,
where $\Delta_0$ labels the magnitude of the parameter
$\Delta ({\bf k})$ and $\Sigma_3$ is the MIR damping energy), 
this  energy should be ascribed  to the AF pseudogap 
($ E_A ({\bf k}_F) - E_S ({\bf k}_F)$,
 with ${\bf k}_F$ on the Fermi surface). 
	In contrast to that, the aforementioned one-band models
had success in explaining this  structure in terms of the parameters
$U/2$ \cite{Rozenberg,Dagotto}, $J$ \cite{Prelovsek} 
or $\Sigma$ \cite{Emery2}.
	The second (charge-transfer) gap appears at 1.5$-$2 eV.
	In the  model A, this energy scale is connected with 
the $pd$ dimerization gap 
($E_L ({\bf k}_F) - E_D ({\bf k}_F)$, $L = N, P$),
and, importantly,  is not simply related to the MIR threshold energy.

	In the gauge invariant formalism \cite{Pines,KupcicPB},
the  optical conductivity of the underdoped high-$T_c$ cuprates 
can be thus shown in the form
\begin{eqnarray}
 \sigma^{\rm total}_{\alpha} ( \omega) &\approx& \frac{{\rm i}}{\omega} 
\frac{e^2 n_{\rm c}^{\rm eff}}{m}
\frac{\hbar \omega}{\hbar \omega + {\rm i} \Sigma_1} 
-{\rm i}  \omega \alpha^{\rm ir}_{\alpha} (\omega, \Sigma_3) 
\nonumber \\
&&
- {\rm i}  \omega  
\alpha^{\rm inter}_{\alpha} (\omega, \Sigma_2) 
- \frac{{\rm i} \omega}{4 \pi}
[\varepsilon_{\alpha,\infty} (\omega) - 1].
\label{eq5} 
\end{eqnarray}
	The four quantities which enter into this expression represent
the effective number of conduction electrons
per unit volume, $n_{\rm c}^{\rm eff}$,
the MIR polarizability, $\alpha^{\rm ir}_{\alpha} (\omega, \Sigma_3)$, 
the interband polarizability, $\alpha^{\rm inter}_{\alpha} (\omega, \Sigma_2)$,
and the contribution of all other, on-site high-frequency optical processes,
$\varepsilon_{\alpha,\infty} (\omega) $.

The $\Sigma_i$ are three phenomenological damping energies.
	As pointed out above, the description of the low-frequency
processes in Eq. (\ref{eq5}) can be easily improved by 
extending the first term to include  
the frequency dependent corrections in the intraband damping energy 
($\Sigma_1 \rightarrow \Sigma_1 (\omega) \equiv \hbar / \tau (\omega )$, 
usually 
$\hbar / \tau (\omega ) = \hbar / \tau + \alpha \hbar \omega$)
and the effective mass (which frequency dependence is forced by the causality
principle, $n_{\rm c}^{\rm eff}/m \rightarrow n/m( \omega)$).
	It should be recalled that the generalized Drude formula 
($\alpha \neq 0$, but $\alpha^{\rm ir}_{\alpha} (\omega) = 0$ and 
$\alpha^{\rm inter}_{\alpha} (\omega)= 0$)	
would  hardly be extended to the underdoped regime
because  large damping energies are required to explain measured spectra 
\cite{Uchida}.
	In the present analysis of the underdoped  regime,
where the MIR optical activity is 
associated  with the excitations across the AF pseudogap, 
rather than with the strong intraband damping effects,
such frequency  corrections to $\Sigma_1$ are presumably small, and will be
disregarded in the quantitative analysis given in Section 3.

	In the model B, the first two terms in Eq. (\ref{eq5}) are given by
\begin{eqnarray}
n_{\rm c}^{\rm eff} &=& \frac{1}{V} \sum_{{\bf k}^* \sigma} 
\gamma^{AA}_{xx} ({\bf k})  [1 -  f_A({\bf k})], 
\label{eq6} \\
\alpha^{\rm ir}_{\alpha } (\omega, \eta) &=& 
\frac{1}{ \omega^2} \frac{1}{V} \sum_{{\bf k}^* \sigma}
\frac{(\hbar \omega)^2 |J_{\alpha}^{AS} ({\bf k})|^2}{
E^2_{AS} ({\bf k}) } 
\nonumber \\ 
&& \times
\frac{2E_{AS} ({\bf k}) [f_A({\bf k}) -  1]}{
(\hbar \omega  + {\rm i} \eta )^2 - E^2_{AS} ({\bf k})  }
.
\label{eq7} \end{eqnarray}
	The sum $\sum_{{\bf k}^*}$ is restricted to the first Brillouin 
zone of the dimerized lattice, and $\gamma^{AA}_{xx} ({\bf k})$ 
is the static Raman vertex defined in Appendix B.
	Similarly, the model A gives rise to the interband polarizability 
of the form
\begin{eqnarray}
\alpha^{\rm inter}_{\alpha } (\omega, \eta) &=& 
\frac{1}{ \omega^2} \frac{1}{V} \sum_{{\bf k} \sigma L \ne D}  
\frac{(\hbar \omega)^2 |J_{\alpha}^{LD} ({\bf k})|^2}{
E^2_{LD} ({\bf k}) }
\nonumber \\ 
&& \times
\frac{- 2E_{LD} ({\bf k})f_D({\bf k})  }{
(\hbar \omega  + {\rm i} \eta )^2 - E^2_{LD} ({\bf k})  }. 
\label{eq8}
\end{eqnarray}
	The Fermi-Dirac function 
$[1 + e^{\beta[E_L({\bf k}) - \mu]}]^{-1}$ is denoted by $f_L({\bf k})$,
and the energy difference 
$E_{L} ({\bf k}) - E_{L'} ({\bf k})$ by $E_{LL'} ({\bf k})$.

Since the contribution of the on-site high-frequency processes
is nearly independent of doping, and
is small at energies  below 3 eV, we take
$Im\{ \varepsilon_{\alpha, \infty} ( \omega) \} = 0$ and 
${ Re} \{ \varepsilon_{\alpha, \infty} ( \omega) \} 
\approx \varepsilon_{\infty}$ in this energy region.
	(A better approximation for
$\varepsilon_{\alpha, \infty} ( \omega)$
is given at the end of the article.) 
	For further considerations it is appropriate to denote the first 
three contributions in Eq. (\ref{eq5})  by
$\sigma^{\rm Drude}_{\alpha} ( \omega)$, $\sigma^{\rm ir}_{\alpha} ( \omega)$
and $\sigma^{\rm inter}_{\alpha} ( \omega)$,  
with the abbreviation 
$\sigma^{\rm intra}_{\alpha} ( \omega) \equiv 
\sigma^{\rm Drude}_{\alpha} ( \omega) + \sigma^{\rm ir}_{\alpha} ( \omega)$.

	The corresponding macroscopic dielectric 
function reads as
\begin{eqnarray}
\varepsilon_{\alpha} (\omega) &=& 1
+ \frac{4 \pi {\rm i}}{\omega }   \sigma^{\rm total}_{\alpha} (\omega),
\label{eq9}\end{eqnarray}
giving finally the three-component optical model with 
the minimal number of adjustable parameters: 
$t_{pd}$, $\Delta_{pd}$, $\Delta_0$,  
$\Sigma_i$ and $\varepsilon_{\infty}$.
	(Actually, we take over the estimate of the ratio
$t_{pd} / \Delta_{pd}$ from the EFG analysis \cite{Kupcicefg},
and reduce additionally the number of independent model parameters.)

\section{Results and discussion}
Let us remember some additional details about the optical anomalies of  
the underdoped and lightly-doped La$_{2-x}$Sr$_x$CuO$_4$ crystals.
	(i) A small transfer of the spectral weight across the 
charge-transfer gap is observed with decreasing doping in the 
$0.2 > \delta > 0.1$ doping region
(the derivative $\partial \Omega_{\rm inter}^2 / \partial \delta$
is negative but  small;
$\Omega_{\rm inter}^2$ is the interband spectral weight to be defined below).
	(ii) As mentioned above, 
this phenomenon is accompanied by the appearance 
of the MIR peak placed at 
$\hbar \omega_{\rm peak}^{\rm ir}  \approx 0.1$ eV.
	(iii) The additional decrease of doping, below $\delta < 0.1$,
is characterized by a dramatic increase of the spectral weight 
above the charge-transfer gap
(i.e. $\partial \Omega_{\rm inter}^2 / \partial \delta < 0$  
becomes  large, resulting for $\delta \approx 0$
in the complete disappearance of the intraband spectral weight).
	(iv) The MIR peak is also shifted, resulting in
$\hbar \omega_{\rm peak}^{\rm ir} \approx 0.15$ eV and  
$\hbar \omega_{\rm peak}^{\rm ir} \approx 0.3$ eV 
at $\delta = 0.1$ and $\delta = 0.06$, respectively.

These features will be discussed now in the framework of the present
three-component optical model.
	The considerations include three important  questions.
	First, starting from the model parameters independent of doping,
we show how the total spectral weight is shared
among  the Drude, MIR  and interband channels.
	Second, we illustrate the dependence of the low-frequency 
optical conductivity on the symmetry and magnitude of  
AF pseudogap.
	Finally, the validity of the model (\ref{eq5})$-$(\ref{eq8})
in the lightly doped regime is tested on comparing  the 
experimental spectra measured at $\hbar \omega < 3$ eV to the 
ones which correspond to the model parameters independent of doping.
 	The characteristic energy scales used in the 
numerical calculations, $\Delta_{pd} = 0.66$ eV and $t_{pd} = 0.73$ eV,	
are deduced from the estimate of the  intracell
hybridization $t_{pd} /\Delta_{pd} \approx 1.1$ \cite{Kupcicefg} 
obtained on the basis of the
intracell charge distributions measured  in the 
La$_{2-x}$Sr$_x$CuO$_4$ compounds by the EFG probes,
and by assuming that the charge-transfer  energy is roughly 
equal to  1.75 eV for the doping level $\delta = 0.2$.

\subsection{Conductivity sum rules}
There has been a lot of interest in the sum rules of 
the high-$T_c$ cuprates, in particular in 
the $c$-axis conductivity sum rule and  the in-plane 
Pines-Nozi\'{e}res  sum rule \cite{Pines,Anderson},
both of which exhibit an indirect,  relatively complicated 
dependence on the MIR threshold  energy.
	Since it is not a trivial task in the three-band model 
to determine these sum rules,
the related theoretical analyses have
employed again the numerical treatments of the one-band models or suitable
phenomenological models \cite{Prelovsek2,Basov,Marsiglio}.
	This contrasts with the direct dependence of 
the in-plane conductivity sum rule of the underdoped compounds
on both the MIR and charge-transfer threshold energy, where it is possible
to make a relatively simple estimation of the  model parameters,
in particular their dependence on the doping level.
	Let us now examine the latter issue in more detail.

The experimental analyses present usually the total spectral weight of 
$\sigma_{\alpha }^{\rm total} ( \omega)$ in terms of the spectral function
$N_{\rm eff} (\hbar \omega)$ defined  by 
\cite{Uchida,Cooper}
\begin{eqnarray}
N_{\rm eff} (\hbar \omega) &=& 
\frac{8}{\Omega_{0}^2} \int^{\omega}_{0}  {\rm d} \omega' \;
{ Re} \{\sigma_{\alpha }^{\rm total} ( \omega') \}. 
\label{eq10}\end{eqnarray}
	The frequency  $\Omega_0 = \sqrt{ 4 \pi e^2 /(mV_{0})}$
is a frequency-scale parameter, and $V_{0}$ is the primitive cell
volume.

	Alternatively, in the multi-component optical models,
several auxiliary frequencies, together with the associated 
effective numbers, can be defined
by considering the spectral weight contained by the $i$-th channel 
\begin{eqnarray}
\Omega_{ i}^2 &=& 4 \int^{\infty}_{-\infty}  {\rm d} \omega \;
{ Re} \{\sigma_{\alpha }^{ i} ( \omega) \} 
\nonumber \\
&\equiv &  V_{0} n^{\rm eff}_{ i} \Omega_{0}^2 .
\label{eq11}\end{eqnarray}
	(To simplify notation, we will refer to 
$\Omega_{\rm i }^2$ as the spectral weight in the $i$-th channel.
	Obviously, this differs from the usual normalization 
of the spectral weight by a factor $1/2$.)
	The index $i \in \{$total, intra, inter, Drude, ir$\}$.
	Notice that $n^{\rm eff}_{\rm Drude}$, defined 
by this relationship, is
identical to $n^{\rm eff}_{\rm c}$ in Eq. (\ref{eq5}).

The conductivity sum rules of the present optical models A and B 
are most appropriately described by
\begin{eqnarray}
\Omega_{\rm total,0 }^2 & = & \Omega_{\rm intra,0  }^2 
+ \Omega_{\rm inter,0 }^2,
\label{eq12} \\
\Omega_{\rm intra }^2 & = & \Omega_{\rm Drude  }^2 + \Omega_{\rm ir  }^2
\label{eq13}\end{eqnarray}
(the label 0 in Eq. (\ref{eq12}) refers to $\Delta ({\bf k}) = 0$).
	For the narrow bands with the relaxation processes
negligible (limit $\Sigma_i \rightarrow 0$), the spectral function 
$N_{\rm eff} (\hbar \omega)$ has a step-like form, 
with the steps simply related to Eqs. (\ref{eq12}) and (\ref{eq13}).
	However, in the typical experimental situations 
(where the bands are wide and the damping energies are not negligible)
the relationship between these two representations of the total spectral weight
is more complicated (see Eqs. (\ref{eq16})). 
	Nevertheless, a careful comparison between the measured 
spectral functions $N_{\rm eff} (\hbar \omega)$
and the analytic expressions for $\Omega_{i }^2$ can be 
used to extract the values of the parameters involved in the 
model (\ref{eq5}), or in some more general optical model
(see Ref. \cite{Tanner}, where the intraband spectral weight and the related
effective number of electrons are estimated in various optimally doped 
cuprates and correlated with $T_c$).

    \begin{figure}[tb]
    \includegraphics[height=18pc,width=18pc]{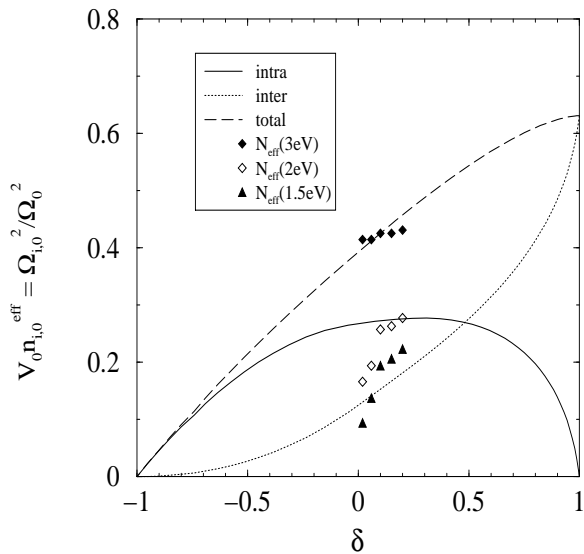}
    \caption{The effective number of electrons
    in the model A as a function of the hole doping.     
   	The experimental points 
	$N_{\rm eff} (1.5 {\rm eV})$ (filled triangles),	
	$N_{\rm eff} (2 {\rm eV})$
    (open diamonds) and $N_{\rm eff} (3 {\rm eV})$ (filled diamonds) 
    measured  in   La$_{2-x}$Sr$_x$CuO$_4$ 
    (Ref. 4, with $\delta = x$ assumed) 
    are also shown.
    }
    \end{figure}

Shown in Fig. 1, the data $N_{\rm eff} (3 {\rm eV})$ measured in 
La$_{2-x}$Sr$_{x}$CuO$_4$ and the total spectral weight of the 
three-band model, 
\begin{eqnarray}
\Omega_{\rm total,0}^2 &=& \Omega_{0}^2\frac{1}{N} \sum_{{\bf k} \sigma} 
(-)\gamma^{DD}_{\alpha \alpha} ({\bf k};2)  f_D ({\bf k}),
\label{eq14}\end{eqnarray}
display the same doping dependence.
	At the same time,  the doping dependence of the intraband 
spectral weight is given by
\begin{eqnarray}
\Omega_{\rm intra,0}^2 &=& \Omega_{0}^2\frac{1}{N} \sum_{{\bf k} \sigma} 
(-)\gamma^{DD}_{\alpha \alpha} ({\bf k})  f_D ({\bf k}). 
\label{eq15}\end{eqnarray}	
	The intra- and interband spectral weights are also shown in Fig.1.
	Interestingly, due to the gauge invariant form of 
$\sigma^{\rm total}_{\alpha} (\omega)$,
none of these three spectral weights  depends on
the damping energies.
	(Here $\gamma^{DD}_{\alpha \alpha} ({\bf k})$ is the 
static Raman vertex of the model A; 
see the effective mass theorem (\ref{eqA3})).

	It should also be recalled that this figure shows the results obtained 
in the simplest case, in which  the parameters $\Delta_{pd}$ and $t_{pd}$ 
are assumed to be independent of doping.
	In this case, for the doping range relevant to the
hole-doped high-$T_c$ cuprates, $0 < \delta < 0.3$, 
the calculated total spectral weight 
is shared between the intra- and interband channels nearly in equal
proportions.

Fit of the predicted spectral weights to the measured data
is based on the relationships
\begin{eqnarray}
\Omega_{\rm total,0}^2 &\approx& 
N_{\rm eff} (\hbar \omega_{\rm max})\Omega_{0}^2 , 
\nonumber \\
\Omega_{\rm intra,0}^2 &\approx& 
N_{\rm eff} (\hbar \omega_{\rm min}) \Omega_{0}^2. 
\label{eq16}\end{eqnarray}
	Here $\hbar \omega_{\rm max}$ and 
$\hbar \omega_{\rm min}$ are  the effective maximal and 
minimal energy in the interband processes, respectively.
	Although there is a large inaccuracy in the experimental 
determination of 
$\hbar \omega_{\rm min}$ and $\hbar \omega_{\rm max}$, 
it seems that $0.2 < N_{\rm eff} (\hbar \omega_{\rm min}) < 0.25$
and $N_{\rm eff} (\hbar \omega_{\rm max}) \approx 0.5$
almost in all La$_2$CuO$_4$ based compounds  \cite{Uchida,Cooper,Quijada}
(see, for example,  Fig. 10 in Ref. \cite{Uchida}).
	In this respect, note
that the MFA of the large $U_d$ case
(i.e. for the average bond energy involved both in the low-
and high-frequency processes),
together with the conclusions of the EFG analysis,
suggests the cutoff energy  $\hbar \omega_{\rm min} \approx 2$ eV 
(Fig. 1), 
contrary to 
the experimental optical studies, which are consistent in the conclusion that 
$\hbar \omega_{\rm min} < 1.5$ eV.
	In Section 3.3 we will turn to a more detailed discussion of this
discrepancy.

In the model B, one obtains that
the spectral weight of the MIR processes
is given by 
\begin{eqnarray}
\Omega_{\rm ir}^2    &=&
\Omega_{0}^2\frac{1}{N} \sum_{{\bf k}^* \sigma} 
\frac{m}{e^2} \frac{2 |J_{\alpha}^{AS} ({\bf k})|^2 }{
 E_{AS} ({\bf k}) }[1- f_A ({\bf k})],  
\label{eq17}\end{eqnarray}
and that of the Drude term by
\begin{eqnarray}
\Omega_{\rm Drude}^2 & =& V_{0} n^{\rm eff}_{\rm c} \Omega_{0}^2.
\label{eq18}\end{eqnarray}
	The resulting  doping dependence of the 
Drude spectral weight is illustrated in Fig. 2, for 
three symmetries of  the effective  potential
\begin{eqnarray}
\Delta ({\bf k}) &=& \Delta_0, 
\nonumber  \\
\Delta ({\bf k}) &=& \frac{1}{2} \Delta_0 (\cos {\bf k} \cdot {\bf a}_{1} 
- \cos {\bf k} \cdot {\bf a}_{2}), 
\nonumber \\ 
\Delta ({\bf k}) &=&  \Delta_0 \sqrt{\frac{1}{2} 
+ \frac{1}{8} (\cos {\bf k} \cdot {\bf a}_{1} 
- \cos {\bf k} \cdot {\bf a}_{2})^2},  
\label{eq19} \end{eqnarray}
corresponding, respectively, to the AF pseudogap of the 
$s$, $d_{x^2-y^2}$ and anisotropic-$s$ symmetry. 
	Again, $t_{pd} = 0.73$ eV is the average bond energy,
and both spectral weights are independent of $\Sigma_i$.

   \begin{figure}[tb]
   \includegraphics[height=18pc,width=18pc]{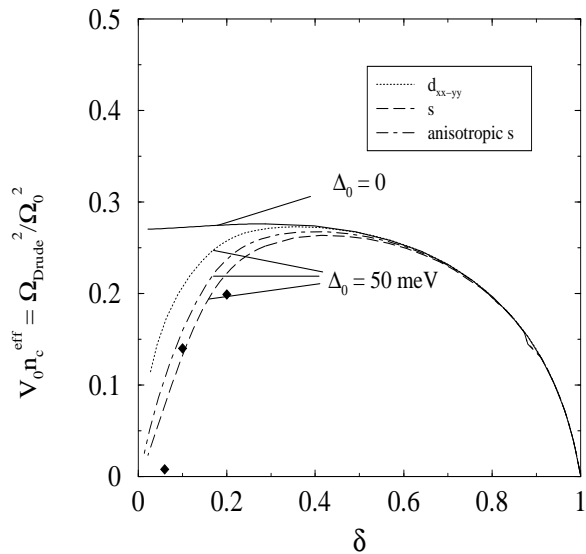}
   \caption{The dependence  of the Drude  spectral weight 
   on the symmetry of the AF pseudogap
    in the hole doped regime, for $\Delta_0 = 50$ meV.
    The filled diamonds represent the square of the measured plasma frequency
    \cite{Uchida} shown in units of 
    $\Omega_0^2/\varepsilon_{\infty}$, as discussed in the text.
    The parameters are
   $\hbar \Omega_0 \approx 3.8$ eV and $\varepsilon_{\infty} \approx 5$.
}
   \end{figure}

The most noticeable feature of the Drude spectral weight  is the occurrence of 
two different  metallic behaviors:
(i) the electron-like behavior characterized by 
$\partial \Omega_{\rm Drude} / \partial \delta < 0$, and 
(ii) the hole-like behavior where 
$\partial \Omega_{\rm Drude} / \partial \delta > 0$.
	 For $\Delta_0 \ll t_{pd}, \Delta_{pd}$, 
the critical doping $\delta_{\rm 0}$, at which 
the behavior of the conduction electrons will be changed,
is only weakly dependent on the AF pseudogap symmetry.
	The critical doping associated with  the curves 
in Fig. 2 agrees well  with
the results of the Hall effect measurements
(where the sign reversal of the Hall coefficient is found at 
$\delta_0 \approx 0.25$ \cite{Uchida2,Uchida}).

	Finally, one can also notice that,  
when the doping level is reduced well below $\delta_0$, 
the effective number of charge carriers obtained 
by the expression (\ref{eq6}) can be adequately described
by the free-hole approximation.
	The effective number  of holes is small with respect to the number of
electrons in the conduction band, $(1 - \delta) /V_{0}$.
	Eq. (\ref{eq18}), together with the definition of the
static Raman vertex, is the relevant expression to reconcile  the
concept of large Fermi surfaces, supported by the ARPES experiments, with
the small effective number of  holes, estimated from the Drude spectral weight
or from the associated electron plasma frequency
(the measured plasma frequencies are indicated in Fig. 2, using 
a simplified single-band expression 
$\Omega_{\rm Drude}^2 \approx 
\varepsilon_{\infty} (\Omega_{\rm pl}^2 + 1/\tau^2) 
\approx \varepsilon_{\infty} \Omega_{\rm pl}^2$).

\subsection{Low-frequency optical conductivity}

According to Fig. 2, the above description of the MIR correlations
becomes ineffective for the   magnitude $\Delta_0$ 
small in comparison with the energy difference between the Fermi energy
and the van Hove energy, $\mu - \varepsilon_{vH}$.
	In such circumstances, the two-component low-frequency optical
conductivity, characterized by the distinct MIR threshold energy,
 will transform into a nearly single-component form
with the increasing role of the frequency dependent corrections 
to the intraband damping energy
\cite{Varma,Ruvalds,Niksic}.
	Not surprisingly, these qualitative changes in the spectra are
correlated with the sign reversal of the Hall coefficient \cite{KupcicSSC}.

   \begin{figure}[tb]
   \includegraphics[height=40pc,width=18pc]{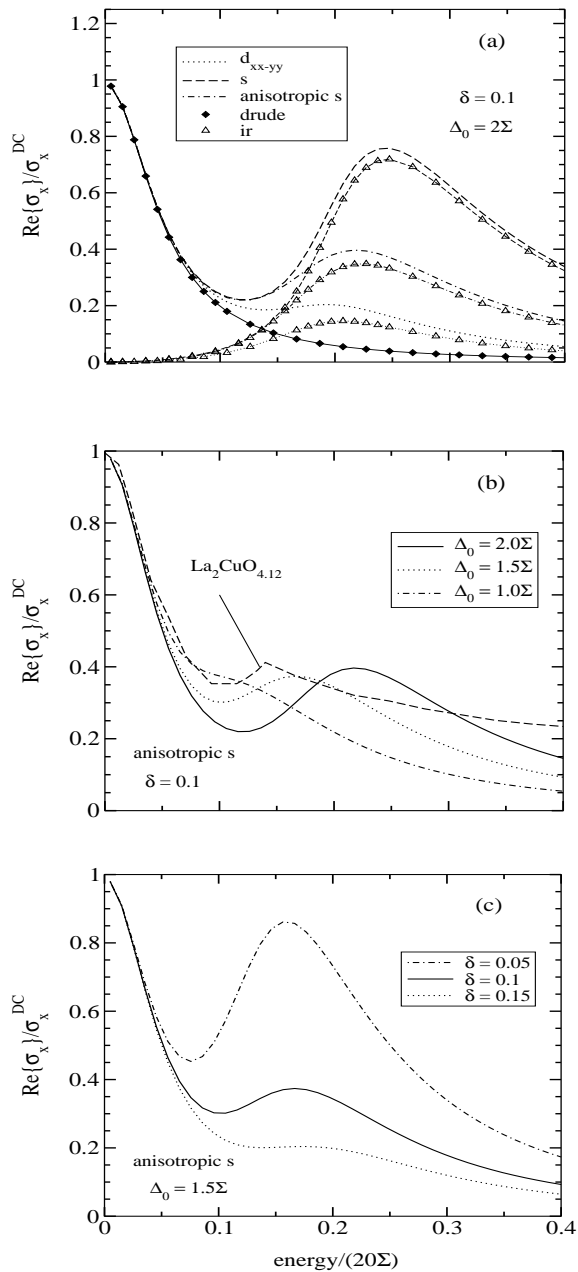}
   \caption{The dependence of the normalized low-frequency
   optical conductivity on the AF pseudogap symmetry (a)
   and magnitude (b), and on the doping level (c),
   $\Sigma_1 = \Sigma_3 = \Sigma$.
	$\sigma^{\rm DC}_{x} = \hbar \Omega_{\rm Drude}^2 /(4 \pi \Sigma)$
   is the DC conductivity.
    	The experimental  data measured in La$_2$CuO$_{4.12}$
    (long-dashed line in Fig. (b)) \cite{Quijada} are given for comparison.
    In this material, $\hbar \omega_{\rm peak}^{\rm ir} 
     \approx 90$ meV and $\Sigma \approx 30$ meV at 
    $T \approx 200$ K.
     }
  \end{figure}	

In order to express the dependence of the low-frequency optical conductivity on
the model parameters more clearly, we show in Fig. 3 the normalized spectra 
for a few characteristic cases.
	First, it should be recalled that the intraband spectral weight is 
directly dependent on the electron group velocity 
$J_{\alpha}^{DD} ({\bf k})/e$, namely
$\Omega_{\rm intra,0}^2 =  V_{0} n^{\rm eff}_{\rm intra} \Omega_{0}^2$
and 
\begin{eqnarray}
n_{\rm intra}^{\rm eff} &=& \frac{m}{e^2} \frac{1}{V} 
\sum_{{\bf k} \sigma} 
[J_{\alpha}^{DD} ({\bf k}) ]^2 \delta [E_D ({\bf k})  - \mu ].
\label{eq20}\end{eqnarray}
	As a result, the transfer of the spectral weight across the 
MIR gap will be  strongly enhanced by the $s$ component of the 
parameter $\Delta ({\bf k})$.
	However, a huge  MIR peak has never been observed 
in La$_{2-x}$Sr$_x$CuO$_{4}$ 
or in La$_2$CuO$_{4+x}$ measured spectra
(for example, compare the relative intensity of the MIR spectrum
measured in La$_2$CuO$_{4.12}$,
the long-dashed line in Fig. 3(b), with the model prediction for 
$\Delta  ({\bf k}) = \Delta_0$, the long-dashed line in Fig. 3(a)).
	Therefore, it seems that $\Delta ({\bf k}) $ is dominantly of the
$d_{x^2-y^2}$ character,
with a small $s$ component directly related to 
the velocity ratio $1/(1 + \lambda)$ measured 
in the ARPES in the nodal ($k_x=k_y$) region of the Fermi surface 
\cite{Lanzara,Valla,Yoshida}.
	Second, in the present model, where the second-neighbor bond-energy
$t_{pp}$ is set to zero, the anomalous shift of the MIR peak structure
with decreasing doping
can only be understood as a result of 
the  doping dependent magnitude $\Delta_0$.
	The lower doping, the stronger effective perturbation has to be,
in full agreement with the conclusion of a similar  single-band analysis
focused on the  characteristic AF  energy, given  in 
Ref. \cite{Friedel}, and with the related specific 
heat measurements \cite{Loram}.
	A significant deviation from such doping dependence of
$\Delta_0$ is expected for $t_{pp}$ comparable to $t_{pd}$,
or for $t_{pp} \approx \Delta_0 > \mu - \varepsilon_{vH}$.
	The difference in the shape of the Fermi surface between
the model prediction for $t_{pp} = 0$ and the experimental
findings can also be removed in this more realistic case.
	The low-frequency optical analysis of this interesting and 
experimentally relevant situation will be given later.

\subsection{Interband optical conductivity}

It is interesting to investigate finally how the expression (\ref{eq5})
relates to the interband optical conductivity data measured 
in the lightly-doped and underdoped compounds.
	Note that both the $U_d = 0$ case and the MFA of the 
large $U_d$ case are unable to explain the anomalous doping dependence
of the interband spectral weight
(compare the model prediction for $\Omega_{\rm inter}^2$ 
with $N_{\rm eff}$(3eV) $- N_{\rm eff}$(1.5eV) in Fig. 1).
	Also, when fitted with $\Delta_{pd} = 0.66$ eV and 
$t_{pd} = 0.73$ eV, the model (\ref{eq5}) underestimates
the interband spectral weight of the considered compounds nearly by a factor 
$N_{\rm eff} {\rm (2eV)} /N_{\rm eff}$(1.5eV).
	The solution of the problem comes on noting that the MFA of the
large $U_d$ case treats the effects of strong local correlations
on the bond energies in an average fashion.
	It is possible, however, to make a simple phenomenological
extension of the present multiband optical model,
and to improve the overall agreement between the model predictions
and the measured data.
	This is due to the gauge invariant form of Eqs.
(\ref{eq5})$-$(\ref{eq8}), which
provides a successful
separation of the twofold role of the intermediate states
in the optical processes.
	For example, the copper-oxygen hopping processes participating in the 
low-frequency optical activity 
are accompanied by $\omega = 0$ and $\Sigma_2 = 0$,
resulting in the effective mass theorem (\ref{eqA3}),
in full agreement with the longitudinal multiband response 
theory \cite{Kupciclong}.
	Similarly, the high-frequency optical 
processes can be described
in terms of the elastic Raman vertices 
(compare the interband polarizability with the difference 
$\gamma^{DD}_{\alpha \alpha} ({\bf k}, \omega) - 
\gamma^{DD}_{\alpha \alpha} ({\bf k})$ in Eq. (\ref{eqA4})),
with the  DC conductivity behaving correctly
in the metal-to-insulator phase transitions \cite{KupcicPB}.

	The parameters describing the low-frequency 
(high-frequency) optical processes in the strongly correlated systems
are (are not) appreciably affected by the strong local correlations.
	The quantity that describes these effects is 
the probability of the photon absorption/emission,
in the low-frequency processes given by the current vertex 
$\tilde{J}^{DD}_{\alpha} ({\bf k})$, in the high-frequency processes
by $\tilde{J}^{DL}_{\alpha} ({\bf k})$. 
	In the strongly correlated regime,
these two vertices should be written in the form  
$\tilde{J}^{DD}_{\alpha} ({\bf k}) \approx t_{pd}^U/t_{pd}
J^{DD}_{\alpha} ({\bf k})$ 
and $\tilde{J}^{DL}_{\alpha} ({\bf k})\approx t_{pd}^0/t_{pd}
J^{DL}_{\alpha} ({\bf k})$,
in the first approximation.
	Here $t_{pd}^0$ is the bare bond energy,
$t_{pd}^U$ is the bond energy renormalized by the large $U_d$, 
$t_{pd}$ is the average energy used in previous sections, and
$J^{LL'}_{\alpha} ({\bf k})$ are given by  the expressions (\ref{eqA1}).
	Consequently, the optical conductivity of the large $U_d$ case
is more appropriately described by
\begin{eqnarray}
\tilde{ \sigma}^{\rm total}_{\alpha} ( \omega) &\approx &
\sum _i \tilde{ \sigma}^{\rm i}_{\alpha} ( \omega), 
\label{eq21}\end{eqnarray}
where
\begin{eqnarray}
\tilde{ \sigma}^{\rm intra}_{\alpha} ( \omega) &\approx &
(t_{pd}^U/t_{pd})^2  \sigma^{\rm intra}_{\alpha} ( \omega), 
\nonumber \\
\tilde{ \sigma}^{\rm inter}_{\alpha} ( \omega) &\approx &
(t_{pd}^0/t_{pd})^2  \sigma^{\rm inter}_{\alpha} ( \omega),
\label{eq22}\end{eqnarray}
or generally 
\begin{eqnarray}
\tilde{ \sigma}^{\rm i}_{\alpha} ( \omega) &\approx &
\zeta_i  \sigma^{\rm i}_{\alpha} ( \omega).
\label{eq23}\end{eqnarray}
	$\zeta_1$, $\zeta_2$, and $\zeta_3$ are three additional
parameters attributed to the Drude, interband and MIR channels,
respectively.

   \begin{figure}[tb]
   \includegraphics[height=18pc,width=18pc]{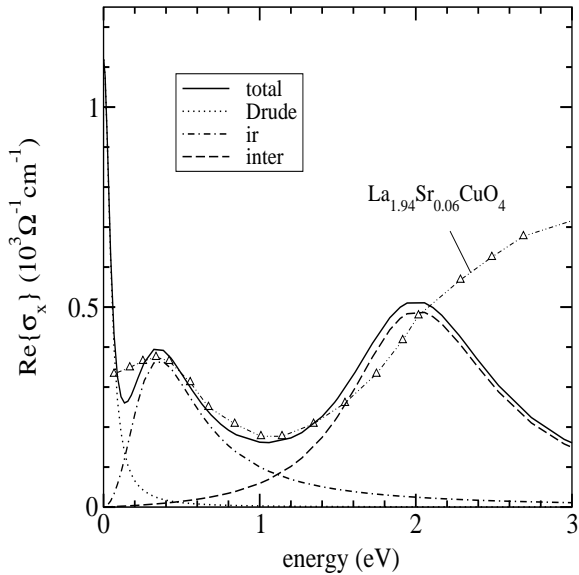}
   \caption{The real part of the optical conductivity for $\delta = 0.06$,
   the $d_{x^2-y^2}$-symmetry AF pseudogap with $\Delta_0 = 0.125$ eV, 
   $\Sigma_1 = 50$ meV, $\Sigma_2 = 0.5$ eV,
   $\Sigma_3 = 0.25$ eV, $\zeta_1 = 0.2$, $\zeta_2 = 1.9$ and $\zeta_3 = 0.8$.
   The experimental data (open triangles) are from Ref. \cite{Uchida}.  }
   \end{figure}

   \begin{figure}[tb]
   \includegraphics[height=18pc,width=18pc]{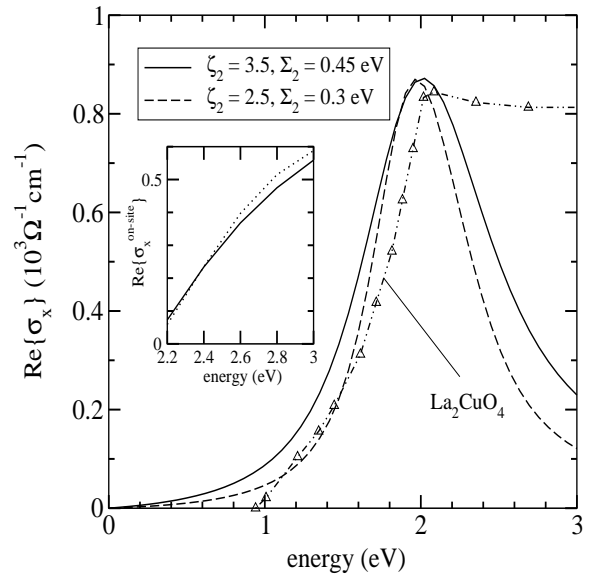}
   \caption{Main figure: 
   The real part of the optical  conductivity 
   at various $\zeta_2$ for $\delta \approx 0$ and $\zeta_1 = \zeta_3 = 0$,
   compared with the experimental data (open triangles) \cite{Uchida}.
   Inset of figure:
   The on-site contributions to the optical conductivity   
    $\omega/(4 \pi) { Im} 
    \{ \varepsilon_{\alpha, \infty} ( \omega) \}$
   (i.e. the three-band-model contributions
   subtracted from the experimental spectra)
    for $\delta = 0.06$ (solid line)  and $\delta = 0$ (dotted
    line).   
     }
   \end{figure}
	
To conclude, due to insufficient information regarding  the low-frequency 
part of the experimental spectra in the lightly-doped crystals,
the discrepancy between the model predictions with
the average bond-energy $t_{pd}$ and the measured data is more clearly 
seen for the interband spectral weight.
	Its anomalous doping dependence 
can be understood as a result of a modification of the bond-energy
renormalizations described by the factor 
$\zeta_2 \approx (t_{pd}^0/t_{pd})^2$ in Eqs. 
(\ref{eq21})$-$(\ref{eq23}).	
	A reasonable agreement of the data measured in the lightly-doped
crystals, shown in Figs. 4 and 5, with the model (\ref{eq23})
is achieved for $\zeta_2 \approx 2$ ($\delta = 0.06$) 
and $\zeta_2 \approx 3$ ($\delta = 0$).
	(For comparison, 
the value $\zeta_2 \approx 1.3$ is obtained by fitting 
the interband spectrum of  La$_{1.8}$Sr$_{0.2}$CuO$_{4}$
to the model (\ref{eq22}).)
	Not surprisingly, similar doping dependence of the bond-energy
renormalizations is found in the previous slave-boson 
analyses \cite{Grilli,Tutis}.
	For example, for the intracell hybridization 
$t^U_{pd}/\Delta_{pd} = 1.1$,
one obtains $(t_{pd}^0/t_{pd}^U)^2 \approx 3.5$ \cite{Tutis},
in a qualitative agreement with the present conclusions.

	Obviously, for the low-frequency part of the spectra, this
is an oversimplified view, because the effects of the strong local 
correlations on these processes are unlikely to be described simply by
multiplying the MFA spectra by the factors 
$\zeta_1 $ and $ \zeta_3$.
	To clarify this essential question, 
further experiments in different high-$T_c$ families are desired as a function 
of the  doping level and with a better resolution of the low-frequency 
part of the spectra. 
	Theoretically, one  needs more accurate investigation
of the relations between the low-frequency optical conductivity 
and the  Raman and ARPES spectra.

Finally, notice that the difference between the experimental 
spectra and the conductivity 
$\tilde{ \sigma}^{\rm total}_{\alpha} ( \omega)$ at energies higher then 
2 eV can be ascribed to the on-site interband optical conductivity
${ Re} \{ \sigma^{\rm on-site}_{\alpha} ( \omega) \}
= \omega/(4 \pi) { Im} 
\{ \varepsilon_{\alpha, \infty} ( \omega) \}$.
	The on-site spectra deduced from the
experimental data and the three-band model spectra
(solid lines in Fig. 4 and 5) are shown in the inset of Fig. 5, 
for the cutoff energy $\hbar \omega_{\rm max} \approx$ 3 eV, and
for $\delta = 0$ and 0.06.
	The nearly identical frequency dependence of 
${ Re} \{ \sigma^{\rm on-site}_{\alpha} ( \omega) \}$ for these two 
doping levels gives an additional, although rough support to
the picture of the interband optical conductivity discussed above.

\section{Conclusion}

In this article, the influence of two dimerization potentials
on the optical response of the valence electrons in the high-$T_c$ cuprates
has been studied, starting from the model parameters independent of doping.
	It is found that for the nearly half filled conduction band 
the $pd$ dimerization potential leads to  
the spectral weights of the intra- and interband channels 
nearly equal to each other.
	It is also shown that at optimum doping 
the effective correlation potential related to the AF pseudogap  processes
results  in the non-Drude nearly single-component 
low-frequency  optical conductivity.
	It is argued that, due to the decrease in the effective
number of conduction electrons, 
the low-frequency conductivity 
transforms naturally from a single-component into a two-component
structure with decreasing doping.
	Both of these results are found to be  in accordance with 
 experimental observation in the lightly-doped and underdoped  
La$_2$CuO$_4$ based compounds. 
	In the present model (where the bond energy $t_{pp}$ is set to zero),
the pronounced doping dependence of the MIR peak structure is attributed
to the doping dependent magnitude of the characteristic AF energy.
	Finally, the substantial increase of the interband spectral weight
 at small doping levels is interpreted in the framework of the conductivity
sum rules as an indirect evidence on the strong renormalizations of the 
parameters describing the low-frequency physics.


\section*{Acknowledgements}

Useful discussions with Professor S. Bari\v{s}i\'{c}
are gratefully acknowledged.
	This work was supported by the Croatian Ministry of Science under the 
project 0119-256.

\appendix

\section{Coupling functions in the model A }
	By using the auxiliary functions 
$u_{\bf k}$, $v_{\bf k}$ and  $t_{\bf k}$ 
defined in Ref. \cite{Kupciclong}, one obtains that
the coupling functions relevant to the case of the nearly half filled bonding
band are
\begin{eqnarray}
J^{DD}_{\alpha} ({\bf k}) &=& 
\frac{e a t^2_{pd}}{\hbar } \frac{2 u_{\bf k} v_{\bf k}}{t_{\bf k}} 
\sin {\bf k} \cdot {\bf a}_{\alpha},
\nonumber  \\
J^{DP}_{\alpha} ({\bf k}) &=& 
 \frac{e a t^2_{pd}}{\hbar }\frac{u^2_{\bf k} -v^2_{\bf k}}{t_{\bf k}} 
\sin {\bf k} \cdot {\bf a}_{\alpha},
\nonumber \\
J^{DN}_{x} ({\bf k}) &=& 
\frac{e a t^2_{pd}}{\hbar } \frac{2u_{\bf k} }{t_{\bf k}} 
\sin \frac{1}{2}{\bf k} \cdot {\bf a}_{2}
\cos \frac{1}{2}{\bf k} \cdot {\bf a}_{1},
\nonumber \\
J^{DN}_{y} ({\bf k}) &=&
- \frac{e a t^2_{pd}}{\hbar }\frac{2u_{\bf k} }{t_{\bf k}} 
 \sin \frac{1}{2}{\bf k} \cdot {\bf a}_{1}
\cos \frac{1}{2}{\bf k} \cdot {\bf a}_{2},
\label{eqA1}\end{eqnarray}
and
\begin{eqnarray}
\gamma ^{DD}_{\alpha \alpha} ({\bf k};2) &=& 
\frac{m}{m_{xx}}
\frac{\Delta_{pd} u_{\bf k} v_{\bf k}}{t_{\bf k}} 
\sin ^2 \frac{1}{2}{\bf k} \cdot {\bf a}_{\alpha}.
\label{eqA2}\end{eqnarray}
	The mass scale has the usual form 
$m_{xx} = \hbar^2 \Delta_{pd}/(2 a^2 t^2_{pd})$, 
with $|{\bf a}_1| = |{\bf a}_2| = a$.

By comparing two expressions for the dimensionless static Raman 
tensor obtained in 
the static limit of the longitudinal \cite{Kupciclong} and 
transverse response of the model A, we can verify 
 the effective mass theorem 
\begin{eqnarray}  
\gamma^{DD} _{\alpha \alpha} ({\bf k}) &=& 
(-) \frac{m}{\hbar^2}
 \frac{\partial^2 E_D ({\bf k}) }{ \partial k_{\alpha}^2} 
\nonumber \\ 
 &=& 
\gamma ^{DD}_{\alpha \alpha} ({\bf k};2 )  
+
\frac{m}{e^2}  
\sum_{L = P, N} 
\frac{2 |J_{\alpha}^{LD} ({\bf k})|^2 }{
 E_{LD} ({\bf k}) } .
 \nonumber \\
\label{eqA3} \end{eqnarray}
	The related inverse {\bf k}-dependent effective mass  
reads as
$$
(1/m) \gamma^{DD} _{\alpha \alpha} ({\bf k}).
$$

    \begin{figure}[tb]
    \includegraphics[height=6pc,width=20pc]{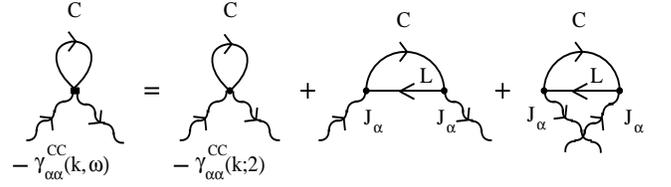}
    \vspace{5mm}
    \caption{    
    The usual diagrammatic representation  of the elastic Raman vertex,
    $L \neq C$.
    In the static limit of the model A (the index $C \equiv D$),
     the summation gives the expression
    (\ref{eqA3}), i.e. the inverse {\bf k}-dependent effective mass.
    The wavy and solid lines represent, respectively,  the photon and 
    electron Green functions.
    Notice  the minus sign in front of the  Raman vertices
    which reflects the fact that electron-like conduction band is shown in the
    hole picture.
   }
    \end{figure}

Beyond the static approximation,
the effective second-order term in the electron-photon coupling
comes from the contributions shown in Fig. 6.
	The elastic Raman  vertex is the vertex function in such
effective processes. 
	The gauge invariant  form of the elastic Raman vertex 
reads as
\begin{eqnarray}
\gamma^{DD} _{\alpha \alpha} ({\bf k}, \omega) &=& 
\gamma^{DD} _{\alpha \alpha} ({\bf k} ) 
-
\frac{m}{e^2}  
\sum_{L = P, N}
 \frac{(\hbar \omega)^2
|J_{\alpha}^{LD} ({\bf k})|^2 }{
E^2_{LD} ({\bf k}) }  
\nonumber \\ 
&& \times
\frac{2E_{LD} ({\bf k})}{
(\hbar \omega  + {\rm i} \eta )^2 - E^2_{LD} ({\bf k})  }. 
\label{eqA4} \end{eqnarray}
	It should be noted that the usual form of the
elastic Raman vertex \cite{Abrikosov,Devereaux1,Sherman}
comes on disregarding  
the factor $(\hbar \omega)^2/ E_{LD}^2 ({\bf k})$ in the expression 
(\ref{eqA4}).

\section{Coupling functions in the model B }
The influence of the perturbation $H_{\rm corr}$ 
on the Bloch functions can be shown
in terms of the auxiliary phase which 
satisfies the relations
\begin{eqnarray}
\cos \varphi_{\bf k} &=& 
\frac{E_D ({\bf k}) - E_{\underline{D}} ({\bf k}) }{E_{AS} ({\bf k})}, 
\nonumber \\
\sin \varphi_{\bf k} &=& 
\frac{ 2\Delta ({\bf k})}{E_{AS} ({\bf k})}.
\label{eqB1}\end{eqnarray}
	The resulting static Raman vertex, the elastic Raman vertex, 
and the current
vertices relevant to the $0 \le \delta \le 1$ doping range are given by
\begin{eqnarray}
\gamma ^{AA}_{\alpha \alpha} ({\bf k}) 
& =& 
\gamma ^{DD}_{\alpha \alpha} ({\bf k}) \cos^2 \frac{\varphi_{\bf k}}{2}
+ \gamma ^{\underline{D}\; \underline{D}}_{\alpha \alpha} ({\bf k} ) 
\sin^2 \frac{\varphi_{\bf k}}{2}
\nonumber \\
&&
 -
\frac{m}{e^2}  
\frac{2 |J_{\alpha}^{AS} ({\bf k})|^2 }{
 E_{AS} ({\bf k}) } 
\nonumber \\
&\approx& 
\gamma ^{DD}_{\alpha \alpha} ({\bf k}) \cos \varphi_{\bf k}
-
\frac{m}{e^2}  
\frac{2 |J_{\alpha}^{AS} ({\bf k})|^2 }{
 E_{AS} ({\bf k}) }, 
 \label{eqB2} \\
\gamma ^{AA}_{\alpha \alpha} ({\bf k}, \omega) &=& 
\gamma^{AA} _{\alpha \alpha} ({\bf k} )  
+ \frac{m}{e^2}  
\frac{(\hbar \omega)^2 |J_{\alpha}^{AS} ({\bf k})|^2}{
E^2_{AS} ({\bf k}) } 
\nonumber \\ 
&& \times
\frac{2E_{AS} ({\bf k})}{
(\hbar \omega  + {\rm i} \eta )^2 - E^2_{AS} ({\bf k})  } , 
\label{eqB3} \end{eqnarray}
and
\begin{eqnarray}
J^{AA}_{\alpha} ({\bf k}) &=& J^{DD}_{\alpha} ({\bf k}) 
\cos^2 \frac{\varphi_{\bf k}}{2}
+J^{\underline{D} \; \underline{D}}_{\alpha} ({\bf k} ) 
\sin^2 \frac{\varphi_{\bf k}}{2}
\nonumber \\
&\approx & J^{DD}_{\alpha} ({\bf k}) 
\cos \varphi_{\bf k}, \nonumber \\
J^{AS}_{\alpha} ({\bf k}) &=& [J^{DD}_{\alpha} ({\bf k}) 
-J^{\underline{D} \; \underline{D}}_{\alpha} ({\bf k} )] 
\sin \frac{\varphi_{\bf k}}{2}\cos \frac{\varphi_{\bf k}}{2} 
\nonumber \\
&\approx&  J^{DD}_{\alpha} ({\bf k}) 
\sin \varphi_{\bf k}. 
\label{eqB4}
\end{eqnarray}
	Again, the labels $\underline{D}$ and $ {\bf k}$ refer to 
the $ {\bf k} \pm {\bf Q}$ states in the $D$ band of the model A,
and 
$\gamma ^{AA}_{\alpha \alpha} ({\bf k};2 ) \approx 
\gamma ^{DD}_{\alpha \alpha} ({\bf k}) \cos \varphi_{\bf k}$
in Eq. (\ref{eq4}).

\end{document}